\documentclass[12pt]{article}
\usepackage{amsfonts}
\usepackage{cite}
\usepackage{amsmath}
\usepackage{amssymb}
\usepackage{graphicx}

\providecommand{\U}[1]{\protect\rule{.1in}{.1in}}
\makeatletter
\@addtoreset{equation}{section}
\makeatother

\input{tcilatex}
\begin{document}

\begin{titlepage}
\begin{center}
\renewcommand{\thefootnote}{\fnsymbol{footnote}}
{\Large{\bf Hypersymplectic Geometry and Supersymmetric Solutions in $(t, s)$ 5D Supergravity}}
\vskip1cm
\vskip 1.3cm
W. A. Sabra
\vskip 1cm
{\small{\it
Department of Physics\\
American University of Beirut\\ Lebanon \\}}
\end{center}
\bigskip
\begin{center}
{\bf Abstract}
\end{center}
Relying on the method of spinorial geometry, purely bosonic supersymmetric solutions in $N=2$, five-dimensional supergravity theories coupled to vector multiplets in all space-time signatures are found. Explicit examples of some new solutions are presented.
\end{titlepage}

\section{Introduction}

The five-dimensional $N=2$ supergravity theories with Lorentzian signature
coupled to vector multiplets was constructed many years ago in \cite{GST}.
In recent years, there has been some interest in supergravity theories in
various space-time signatures. The Euclidean versions of the supergravity
theories of \cite{GST} were considered in \cite{e5}, where it was
demonstrated that the Lagrangian of the Euclidean theory has the kinetic
terms of the gauge fields with the non-conventional sign. Euclidean $N=2$
theories in four dimensions were first considered in \cite{mohaupt1,
mohaupt2, mohaupt3, mohaupt4}. The Euclidean four-dimensional $N=2$
supergravity theories were obtained as dimensional reductions of $N=2$, $D=5$
supergravity theories on a time-like circle. The reduction of the
five-dimensional Euclidean theory on a circle produces Euclidean $N=2$
four-dimensional supergravity with the non-conventional signs of the gauge
fields kinetic terms. The four-dimensional Euclidean supergravity theory
with vector and hypermultiplets was also be obtained via the dimensional
reduction of Euclidean ten-dimensional type IIA supergravity over a
Calabi-Yau threefold, $CY_{3}$ \cite{tenred}. A class of Lorentzian
five-dimensional $N=2$ supergravity theories constructed in \cite{GST} is
obtainable via the dimensional reduction of the standard eleven-dimensional
supergravity \cite{cjs} on a $CY_{3}$ \cite{cad}$.$ Recently, in \cite%
{special}, $N=2$ four and five-dimensional supergravity theories in
space-time signatures $\left( t,s\right) ,$ where $t$ and $s$ are
respectively the number of time and spatial dimensions, were constructed by
reducing Hull's eleven-dimensional supergravity \cite{Hull} on $CY_{3}$. For
a detailed analysis on supersymmetry algebras in arbitrary space-time
dimension and signature we refer the reader to \cite{thesis}.

The eleven dimensional supergravity theories of Hull with space-time
signatures $\left( 1,10\right) $, $\left( 5,6\right) $ and ($9,2)$ have
actions with the standard conventional sign for the 3-form gauge kinetic
term. The mirror theories with signatures $\left( 10,1\right) $, $\left(
6,5\right) $ and $\left( 2,9\right) $ all have the non-conventional sign for
the 3-form gauge fields kinetic terms. In the reduction of the theories with
signatures $\left( 1,10\right) $, $\left( 5,6\right) $ and $\left(
2,9\right) ,$ the $CY_{3}$ is taken to be of signature $\left( 0,6\right) $.
For the reduction of theories with signatures $\left( 10,1\right) $, $\left(
6,5\right) $ and $\left( 9,2\right) ,$ the $CY_{3}$ is of signature $\left(
6,0\right) $.

By employing the methods of \cite{gaun}, a systematic classification of
supersymmetric solutions of the $(1,4)$ five-dimensional minimal
supergravity was given in \cite{original}. In this approach, the existence
of at least one Killing spinor is assumed and differential forms as
bilinears in terms of this spinor are constructed. The algebraic and
differential constraints satisfied by the bilinears can be used to fix the
solution of the space-time metric in addition to the bosonic fields of the
supersymmetric solution. It was found in \cite{original} that
half-supersymmetric solutions with time-like Killing vectors have a
four-dimensional base space given by a hyper-K\"{a}hler manifold. These
findings for the time-like solutions were generalised to supergravity
theories coupled to arbitrary many abelian vector multiplets in \cite{unique}
where also a uniqueness theorem for asymptotically flat supersymmetric black
holes with regular horizons was given.

The goal of our present work is the generalisation of the results \cite%
{unique} to all $N=2$, five-dimensional supergravity theories coupled to
vector multiplets in all space-time signatures. The Killing spinor equations
shall be analysed using the spinorial geometry methods which were first
employed in the analysis of supersymmetric solutions in ten and eleven
dimensions\ in \cite{spin}. Spinorial geometry \cite{geometry} has been very
useful and efficient in the classifications of solutions with various
fractions of supersymmetry in all space-time dimensions \cite{report}.

We organise our work as follows. In section 2, a review of some of the basic
properties of the ungauged five-dimensional supergravity coupled to
arbitrary many vector multiplets is given. Section three contains the
analysis of supersymmetric solutions where the set of rules for the
construction of these solutions is given. Some examples and a summary are
given in section 4.

\section{$(t,s)$ Five-Dimensional Supergravity}

Ignoring hypermultiplets, the bosonic action of the theory for all $N=2$, $%
D=5$ supergravity contains the gravity multiplet and vector multiplets and
is given by \cite{GST, special} 
\begin{equation}
S_{5}=\int_{M_{5}}\frac{1}{2}R\hat{\ast}1-\frac{1}{2}Q_{IJ}(X)dX^{I}\wedge
\ast dX^{J}+\frac{\kappa ^{2}}{4}Q_{IJ}(X)F^{I}\wedge \hat{\ast}F^{J}-\frac{1%
}{12}C_{IJK}A^{I}\wedge F^{J}\wedge F^{K}  \label{action}
\end{equation}%
where $C_{IJK}$ are real constants symmetric in $I,J,K.$ We have $\kappa
^{2}=-1,$ for signature $\left( 1,4\right) $, $\left( 5,0\right) $ and $%
\left( 3,2\right) $ theories and $\kappa ^{2}=1$ for signature $\left(
4,1\right) $,$\left( 0,5\right) $ and $\left( 2,3\right) .$ Here $F^{I}$ are
two-forms representing the gauge fields. The information about the theory is
encoded in the cubic prepotential which describes very special geometry 
\begin{equation}
\mathcal{V}={\frac{1}{6}}C_{IJK}X^{I}X^{J}X^{K}\,=1,
\end{equation}%
$X^{I}$ being the very special coordinates, functions of the $n$ real scalar
fields belonging to the vector multiplets.

The gauge coupling metric can be derived from the prepotential and is given 
\begin{equation}
Q_{IJ}=-{\frac{1}{2}}\left( {\partial _{X^{I}}}{\partial _{X^{J}}}(\ln 
\mathcal{V})\right) _{\mathcal{V}=1}=\frac{1}{2}\left( {9}%
X_{I}X_{J}-C_{IJK}X^{K}\right) ,  \label{ga}
\end{equation}%
where the dual fields $X_{I}$ are defined by 
\begin{equation}
X_{I}={\frac{1}{6}}C_{IJK}X^{I}X^{K}.  \label{d}
\end{equation}%
We also have the useful relations

\begin{equation}
Q_{IJ}X^{J}={\frac{3}{2}}X_{I}\,,\qquad Q_{IJ}dX^{J}=-{\frac{3}{2}}dX_{I}\,.
\end{equation}%
The Killing spinor equations associated with the above theories are given 
\begin{equation}
\left[ \nabla _{\mu }+\frac{\kappa }{8}H_{\rho \sigma }\left( \Gamma _{\mu
}\Gamma ^{\rho \sigma }-6\delta _{\mu }^{\rho }\Gamma ^{\sigma }\right) %
\right] \epsilon =0  \label{one}
\end{equation}%
and 
\begin{equation}
\left( \kappa G_{\mu \nu }^{I}\Gamma ^{\mu \nu }-2\partial _{\mu
}X^{I}\Gamma ^{\mu }\right) \epsilon =0,  \label{two}
\end{equation}
where 
\begin{eqnarray}
G_{\mu \nu }^{I} &=&F_{\mu \nu }^{I}-X^{I}X_{J}F_{\mu \nu }^{J},  \notag \\
\nabla _{\mu } &=&\partial _{\mu }+\frac{1}{4}\omega _{\mu ,\rho \sigma
}\Gamma ^{\rho \sigma },  \notag \\
H_{\mu \nu } &=&X_{I}F_{\mu \nu }^{I}.
\end{eqnarray}

Here $\Gamma _{\mu }$ are Dirac matrices and $\omega _{\mu ,\rho \sigma }$
are the spin connections. For the supergravity theories with space-time
signature $(1,4)$, $(3,2)$ and $(5,0)$, we have $\kappa =-i.$ For the
supergravity theories with space-time signatures $(4,1),(2,3)$ and $(0,5)$,
we have $\kappa =1$.

\section{Supersymmetric solutions}

In what follows, we find solutions admitting Killing spinors through the
analysis of the Killing spinor equations (\ref{one}) and (\ref{two}) using
spinorial geometry methods. We take the Dirac spinors to be the space of
complex forms on $\mathbb{R}^{2}$ spanned over $\mathbb{C}$ by $1,$ $e_{1},$ 
$e_{2}$ and $e_{12}=e_{1}$ $\wedge $ $e_{2}$. \ To proceed in the analysis
of solutions admitting Killing spinors, we start by writing our metric
solutions in the form 
\begin{eqnarray}
ds_{5}^{2} &=&\kappa ^{2}\left( \mathbf{e}^{5}\right) ^{2}+\eta _{\alpha 
\bar{\beta}}\mathbf{e}^{\alpha }\mathbf{e}^{\bar{\beta}}  \notag \\
&=&\kappa ^{2}\left( \mathbf{e}^{5}\right) ^{2}+2\left( \kappa _{1}^{2}%
\mathbf{e}^{1}\mathbf{e}^{\bar{1}}+\kappa _{2}^{2}\mathbf{e}^{2}\mathbf{e}^{%
\bar{2}}\right)
\end{eqnarray}
where $\kappa ^{2},$ $\kappa _{1}^{2}$ and $\kappa _{2}^{2}$\textbf{\ }are
chosen to be $\pm 1,$ depending on the space-time signature of the
considered theory. For example, if we are considering the supergravity
theories with $(2,3)$ signature, we take $\kappa ^{2}=\kappa _{1}^{2}=1,$ $%
\kappa _{2}^{2}=-1$ or alternatively $\kappa ^{2}=\kappa _{2}^{2}=1,$ $%
\kappa _{1}^{2}=-1.$

The action of the $\Gamma $-matrices on spinors is given by 
\begin{eqnarray*}
\Gamma _{1} &=&\kappa _{1}\sqrt{2}e^{1}\wedge ,\ \ \ \ \ \Gamma _{\bar{1}%
}=\kappa _{1}\sqrt{2}i_{e^{1}},\ \ \ \Gamma _{2}=\sqrt{2}\kappa
_{2}e^{2}\wedge ,\ \ \ \ \Gamma _{\bar{2}}=\sqrt{2}\kappa _{2}i_{e^{2}},\  \\
\Gamma _{5}1 &=&\kappa 1,\ \ \ \ \ \ \ \ \ \ \ \ \ \ \ \ \ \Gamma
_{5}e_{1}=-\kappa e_{1},\ \ \ \ \ \Gamma _{5}e_{2}=-\kappa e_{2},\ \ \ \ \ \
\ \ \Gamma _{5}e_{12}=\kappa e_{12}.
\end{eqnarray*}%
We shall find solutions for the Killing spinor $\epsilon =f1.$ This Killing
spinor orbit corresponds to time-like solutions in the standard supergravity
models with signature $(1,4)$. Plugging $\epsilon =f1$ in the Killing spinor
equation (\ref{one}), we obtain the following conditions 
\begin{eqnarray}
\partial _{\alpha }\log f+\frac{1}{2}\omega _{\alpha ,\mu }^{\text{ \ \ \ \
\ }\mu }-\frac{3}{4}H_{\alpha 5} &=&0,\text{ \ \ \ \ \ }  \notag \\
\partial _{\bar{\alpha}}\log f+\frac{1}{2}\omega _{\bar{\alpha},\mu }^{\text{
\ \ \ \ \ }\mu }-\frac{1}{4}H_{\bar{\alpha}5} &=&0,  \notag \\
\partial _{5}\log f+\frac{1}{2}\omega _{5,\mu }^{\text{ \ \ \ \ \ }\mu }+%
\frac{1}{4}\kappa ^{2}H_{\mu }^{\text{ \ \ }\mu } &=&0,\text{ \ \ \ \ \ } 
\notag \\
\kappa ^{2}\omega _{1,\bar{1}5}+\frac{1}{2}\kappa _{1}^{2}H_{\mu }^{\text{ \
\ }\mu }-\frac{3}{2}H_{1\bar{1}} &=&0,  \notag \\
\kappa ^{2}\omega _{2,\bar{2}5}+\frac{1}{2}\kappa _{2}^{2}H_{\mu }^{\text{ \
\ }\mu }-\frac{3}{2}H_{2\bar{2}} &=&0,\text{ \ \ \ \ \ }  \notag \\
\kappa ^{2}\omega _{2,\bar{1}5}-\frac{3}{2}H_{2\bar{1}} &=&0,  \notag \\
\kappa ^{2}\omega _{1,\bar{2}5}\text{\ }-\frac{3}{2}H_{1\bar{2}} &=&0,\text{
\ \ \ \ \ }  \notag \\
\omega _{\bar{2},\bar{1}5}-\frac{1}{2}\kappa ^{2}H_{\bar{2}\bar{1}} &=&0, 
\notag \\
\omega _{\bar{1},\bar{2}5}-\frac{\kappa ^{2}}{2}H_{\bar{1}\bar{2}} &=&0,%
\text{ \ \ \ \ \ }  \notag \\
\omega _{1,\bar{1}\bar{2}}+\frac{1}{2}\kappa _{1}^{2}H_{\bar{2}5} &=&0, 
\notag \\
\omega _{2,\bar{1}\bar{2}}-\frac{1}{2}\kappa _{2}^{2}H_{\bar{1}5} &=&0,\text{
\ \ \ \ \ }  \notag \\
\omega _{\alpha ,\beta \gamma } &=&0  \notag \\
\omega _{\alpha ,\alpha 5} &=&0,  \notag \\
\omega _{5,\bar{1}5}-\kappa ^{2}H_{5\bar{1}} &=&0,\text{ \ \ \ \ \ }  \notag
\\
\omega _{5,\bar{2}5}-\kappa ^{2}H_{5\bar{2}} &=&0,\text{ \ \ \ \ \ }  \notag
\\
\omega _{5,\bar{1}\bar{2}}+\frac{1}{2}\kappa ^{2}H_{\bar{1}\bar{2}} &=&0.
\end{eqnarray}%
The analysis of this linear system of equations implies the following
conditions 
\begin{eqnarray}
\partial _{5}f &=&0,\text{ \ \ \ }  \notag \\
\omega _{5,5\alpha }-2\kappa ^{2}\partial _{\alpha }\log f &=&0,  \notag \\
\omega _{5,\alpha \beta }+\omega _{\alpha ,\beta 5} &=&0,  \notag \\
\omega _{\bar{\beta},\alpha 5}+\omega _{\alpha ,\bar{\beta}5} &=&0,\text{\ \
\ }  \notag \\
\text{\ }\omega _{5,\mu }^{\text{ \ \ \ \ \ }\mu }-\omega _{\mu ,5}^{\text{
\ \ \ \ \ }\mu } &=&0,  \notag \\
\omega _{\alpha ,\mu }^{\text{ \ \ \ \ \ }\mu }-\partial _{\alpha }\log f
&=&0,\text{\ \ }  \notag \\
\omega _{\alpha ,\beta \gamma } &=&0,  \notag \\
\omega _{\alpha ,\bar{\mu}\bar{\nu}}+\eta _{\alpha \bar{\mu}}\partial _{\bar{%
\nu}}\log f-\eta _{\alpha \bar{\nu}}\partial _{\bar{\mu}}\log f &=&0,
\label{con}
\end{eqnarray}%
and 
\begin{eqnarray}
H_{\alpha \beta } &=&-2\kappa ^{2}\omega _{5,\alpha \beta },  \notag \\
\text{\ }H_{\alpha 5} &=&2\partial _{\alpha }\log f,  \notag \\
H_{\alpha \bar{\beta}} &=&\frac{2}{3}\kappa ^{2}\left( \omega _{\alpha ,\bar{%
\beta}5}-\eta _{\alpha \bar{\beta}}\omega _{5,\mu }^{\text{ \ \ \ \ \ }\mu
}\right) ,  \notag \\
H_{\mu }^{\text{ \ \ }\mu } &=&-2\kappa ^{2}\omega _{5,\mu }^{\text{ \ \ \ \
\ }\mu }.  \label{gc}
\end{eqnarray}%
The analysis of (\ref{two}) gives the conditions 
\begin{eqnarray}
F_{\mu }^{I\text{ \ }\mu } &=&X^{I}H_{\mu }^{\text{ \ }\mu }\text{\ }, 
\notag \\
F_{5\alpha }^{I\text{ \ }} &=&X^{I}H_{5\alpha }^{\text{ \ }}\text{\ }-{%
\partial }_{\alpha }X^{I},  \notag \\
F_{\alpha \beta }^{I\text{ \ }} &=&X^{I}H_{\alpha \beta }^{\text{ \ }}\text{%
\ },  \notag \\
{\partial }_{5}X^{I} &=&0.  \label{se}
\end{eqnarray}%
To proceed, we define the 1-form 
\begin{equation}
V=f^{2}\mathbf{e}^{5},
\end{equation}%
and introduce the coordinate $\tau $ such that the dual vector field is
given by $\kappa ^{2}f^{2}\frac{\partial }{\partial \tau }.$ The first four
conditions in (\ref{con}) provide the necessary and sufficient conditions
for $V$ to define a Killing vector. Also those conditions imply that 
\begin{equation}
\mathcal{L}_{V}\mathbf{e}^{5}=0.
\end{equation}%
Furthermore one finds 
\begin{eqnarray}
\mathcal{L}_{V}\mathbf{e}^{1} &=&-\kappa _{1}^{2}f^{2}\left[ \left( \omega
_{5,\bar{1}1}-\omega _{1,\bar{1}5}\right) \mathbf{e}^{1}+\left( \omega _{5,%
\bar{1}2}-\omega _{2,\bar{1}5}\right) \mathbf{e}^{2}\right]   \notag \\
\mathcal{L}_{V}\mathbf{e}^{2} &=&-\kappa _{2}^{2}f^{2}\left[ \left( \omega
_{5,\bar{2}1}-\omega _{1,\bar{2}5}\right) \mathbf{e}^{1}+\left( \omega _{5,%
\bar{2}2}-\omega _{2,\bar{2}5}\right) \mathbf{e}^{2}\right] 
\end{eqnarray}%
By making an appropriate gauge transformation as discussed in \cite{ex}, we
can set 
\begin{equation}
\text{\ \ }\mathcal{L}_{V}\mathbf{e}^{\alpha }=0.
\end{equation}%
We can choose coordinates such that 
\begin{equation}
\mathbf{e}^{5}=f^{2}\left( d\tau +w\right) ,\text{ \ \ \ \ \ }\mathbf{e}%
^{\alpha }=f^{-1}\mathbf{E}^{\alpha }.
\end{equation}%
where the function $f$, the one-form $w$ and $\mathbf{E}^{\alpha }$ are all
independent of the coordinate $\tau .$ At this stage, we define the
following three two-forms: 
\begin{align}
\mathbf{J}_{1}& =\mathbf{E}^{1}\wedge \mathbf{E}^{2}+\mathbf{E}^{\bar{1}%
}\wedge \mathbf{E}^{\bar{2}},  \notag \\
\mathbf{J}_{2}& =-i\left( \mathbf{E}^{1}\wedge \mathbf{E}^{2}-\mathbf{E}^{%
\bar{1}}\wedge \mathbf{E}^{\bar{2}}\right) ,  \notag \\
\mathbf{J}_{3}& =i\left( \kappa _{1}^{2}\mathbf{E}^{1}\wedge \mathbf{E}^{%
\bar{1}}+\kappa _{2}^{2}\mathbf{E}^{2}\wedge \mathbf{E}^{\bar{2}}\right) .
\end{align}%
It can be shown that 
\begin{equation*}
d\mathbf{J}_{1}=d\mathbf{J}_{2}=d\mathbf{J}_{3}=0
\end{equation*}%
provided 
\begin{equation}
\Omega _{\alpha ,\bar{\mu}\bar{\nu}}=0,\text{ \ \ \ \ }\Omega _{\alpha
,\beta \gamma }=0,\text{ \ \ \ }\Omega _{\alpha ,\mu }^{\text{ \ \ \ \ \ \ }%
\mu }=0,  \label{cov}
\end{equation}%
where $\Omega $ represent the spin connections of the base manifold with
vielbeins $\mathbf{E}^{\alpha }.$ In fact, the conditions in (\ref{cov}) are
implied by the last three conditions of (\ref{con}). Moreover, $\mathbf{J}%
_{i},$ $i=1,2,3,$ are covariantly constant two-forms on the base manifold.
They also satisfy the following algebra 
\begin{equation}
\mathbf{J}_{1}^{2}=\mathbf{J}_{2}^{2}=-\kappa _{1}^{2}\kappa _{2}^{2},\text{
\ \ \ \ }\mathbf{J}_{3}^{2}=-1\mathbf{,}\text{ \ \ \ \ }\mathbf{J}_{1}%
\mathbf{J}_{2}=-\mathbf{J}_{2}\mathbf{J}_{1}=-\kappa _{1}^{2}\kappa _{2}^{2}%
\mathbf{J}_{3}.  \label{algebra}
\end{equation}

For $\kappa _{1}^{2}\kappa _{2}^{2}=-1,$ relevant for theories with
space-time signatures $(2,3)$ and $(3,2),$ the algebra (\ref{algebra}) is
that of para-quaternions or the so-called split quaternions \cite{algebra}.
We shall refer to the base manifold with such a structure as hypersymplectic 
\cite{hitch}. For the cases with $\kappa _{1}^{2}=\kappa _{2}^{2}=\pm 1,$
relevant for space-time signatures $(1,4),$ $(4,1)$, $(5,0)$ and $(0,5),$
the algebra (\ref{algebra}) defines the algebra of quaternions.

We now turn back to the analysis of the gauge fields. Using (\ref{gc}), we
have 
\begin{equation}
H=-2\kappa ^{2}\omega _{5,\alpha \beta }\mathbf{e}^{\alpha }\wedge \mathbf{e}%
^{\beta }\text{\ }+2\partial _{\alpha }\log f\mathbf{e}^{\alpha }\wedge 
\mathbf{e}^{5}+\frac{2}{3}\kappa ^{2}\left( \omega _{\alpha ,\bar{\beta}%
5}-\eta _{\alpha \bar{\beta}}\omega _{5,\mu }^{\text{ \ \ \ \ \ }\mu
}\right) \mathbf{e}^{\alpha }\wedge \mathbf{e}^{\bar{\beta}}.
\end{equation}%
Noting that 
\begin{eqnarray}
&&\kappa ^{2}d\mathbf{e}^{5}=-2\kappa ^{2}\mathbf{e}^{5}\wedge d\log
f+2\left( \omega _{1,5\bar{1}}\mathbf{e}^{\bar{1}}\wedge \mathbf{e}%
^{1}+\omega _{2,5\bar{2}}\mathbf{e}^{\bar{2}}\wedge \mathbf{e}^{2}\right)  \\
&&+2\left( \omega _{\bar{1},52}\mathbf{e}^{2}+\omega _{\bar{1},5\bar{2}}%
\mathbf{e}^{\bar{2}}\right) \wedge \mathbf{e}^{\bar{1}}+2\left( \omega
_{1,52}\mathbf{e}^{2}+\omega _{1,5\bar{2}}\mathbf{e}^{\bar{2}}\right) \wedge 
\mathbf{e}^{1}.  \notag
\end{eqnarray}%
We obtain 
\begin{eqnarray}
H-d\mathbf{e}^{5} &=&-\frac{2}{3}\kappa ^{2}\left( \omega _{1,5\bar{1}}%
\mathbf{e}^{\bar{1}}\wedge \mathbf{e}^{1}+\omega _{2,5\bar{2}}\mathbf{e}^{%
\bar{2}}\wedge \mathbf{e}^{2}+2\omega _{1,5\bar{2}}\mathbf{e}^{\bar{2}%
}\wedge \mathbf{e}^{1}+2\omega _{\bar{1},52}\mathbf{e}^{2}\wedge \mathbf{e}^{%
\bar{1}}\right)   \notag \\
&&-\frac{2}{3}\kappa ^{2}\kappa _{1}^{2}\kappa _{2}^{2}\left( \omega _{5,2%
\bar{2}}\mathbf{e}^{1}\wedge \mathbf{e}^{\bar{1}}+\omega _{5,1\bar{1}}^{%
\text{ \ \ \ \ }}\mathbf{e}^{2}\wedge \mathbf{e}^{\bar{2}}\right) .
\end{eqnarray}%
The right hand of the above equation can be expressed in terms of the self
dual part of $dw,$ and we have 
\begin{equation}
H=d\mathbf{e}^{5}-\frac{f^{2}}{3}\left( dw+\ast dw\right) .
\end{equation}%
where our orientation is such that $\epsilon _{1\bar{1}2\bar{2}}=\kappa
_{1}^{2}\kappa _{2}^{2}.$ If we write%
\begin{equation}
f^{2}dw=G_{+}+G_{-},  \label{s}
\end{equation}%
then we have 
\begin{equation*}
H=d\mathbf{e}^{5}+\Psi 
\end{equation*}%
with 
\begin{equation}
\Psi =-\frac{2}{3}G_{+}  \label{l}
\end{equation}%
and thus $\Psi $ is a self-dual 2-form on the base manifold. Using (\ref{se}%
), we find 
\begin{equation}
F^{I}=d\left( X^{I}{\mathbf{e}}^{5}\right) +\Psi ^{I}  \label{g}
\end{equation}%
where $\Psi =X_{I}\Psi ^{I}.$ The Bianchi identity then implies 
\begin{equation}
d\Psi ^{I}=0  \label{t}
\end{equation}%
and thus $\Psi ^{I}$ are harmonic self-dual 2-forms on the base manifold
with metric $ds_{4}^{2}=\eta _{\alpha \bar{\beta}}\mathbf{e}^{\alpha }%
\mathbf{e}^{\bar{\beta}}.$ Turning to Maxwell equations 
\begin{equation}
d(Q_{IJ}\ast F^{J})=\frac{\kappa ^{2}}{4}C_{IJK}F^{J}\wedge F^{K}
\end{equation}%
we obtain after some calculation 
\begin{equation}
\nabla ^{2}\left( f^{-2}X_{I}\right) =-\frac{\kappa ^{2}}{6}C_{IJK}\Psi
^{J}.\Psi ^{K}  \label{ge}
\end{equation}%
where the Laplacian is for the metric $ds_{4}^{2}=\eta _{\alpha \bar{\beta}}%
\mathbf{E}^{\alpha }\mathbf{E}^{\bar{\beta}}$ and we have the convention
that for two $p$-forms $\alpha $ and $\beta ,$ we have 
\begin{equation}
\alpha .\beta =\frac{1}{p!}\alpha _{n_{1}...n_{2}}\beta ^{n_{1}...n_{2}}.
\end{equation}%
Finally we note that the integrability conditions for the Killing spinor
equations together with imposing the Bianchi identity and the equations of
motion for the gauge fields guarantee that all the equations of motion are
satisfied. 

\subsection{Examples and discussion}

In \cite{original} solutions for the minimal case (no vector multiplets)
with base space $\mathbb{R}^{4}$ were constructed. The four dimensional base
metric can be expressed in terms of the left or right invariant forms of $%
SU(2)$ given in terms of Euler angles$.$ The right invariant one forms are
given by 
\begin{equation}
\sigma _{1}=\sin \phi d\theta -\cos \phi \sin \theta d\psi ,\text{ \ \ \ \ }%
\sigma _{2}=\cos \phi d\theta +\sin \theta \sin \phi d\psi ,\text{ \ \ \ \ }%
\sigma _{3}=d\phi +\cos \theta d\psi 
\end{equation}%
and the left invariant ones are given by 
\begin{equation}
\chi _{1}=-\sin \psi d\theta +\cos \psi \sin \theta d\phi ,\text{ \ \ \ \ }%
\chi _{2}=\cos \psi d\theta +\sin \theta \sin \psi d\phi ,\text{ \ \ \ \ }%
\chi _{3}=d\psi +\cos \theta d\phi 
\end{equation}%
satisfying 
\begin{equation}
d\sigma _{i}=-\frac{1}{2}\epsilon _{ijk}\sigma _{j}\wedge \sigma _{k},\text{
\ \ \ \ }d\chi _{i}=\frac{1}{2}\epsilon _{ijk}\chi _{j}\wedge \chi _{k}.
\end{equation}%
In terms of these forms, the flat four-dimensional metric can be written in
the form%
\begin{eqnarray}
ds_{4}^{2} &=&dr^{2}+\frac{r^{2}}{4}\left( \sigma _{1}^{2}+\sigma
_{2}^{2}+\sigma _{3}^{2}\right)   \notag \\
&=&dr^{2}+\frac{r^{2}}{4}\left( \chi _{1}^{2}+\chi _{2}^{2}+\chi
_{3}^{2}\right)   \notag \\
&=&dr^{2}+\frac{r^{2}}{4}\left( d\theta ^{2}+\sin ^{2}d\phi ^{2}+\left(
d\psi +\cos \theta d\phi \right) ^{2}\right) .
\end{eqnarray}%
Defining%
\begin{equation}
\mathbf{E}^{1}=\frac{1}{\sqrt{2}}\left( e^{0}+ie^{3}\right) \text{\ \ \ \ }%
\mathbf{E}^{2}=\frac{1}{\sqrt{2}}\left( e^{2}+ie^{1}\right) 
\end{equation}%
with 
\begin{equation}
e^{0}=dr,\text{ \ \ \ }e^{1}=\frac{1}{2}r\sigma _{1},\text{ \ \ \ }e^{2}=%
\frac{1}{2}r\sigma _{2}\text{ \ \ \ \ \ }e^{3}=\frac{1}{2}r\sigma _{3},
\end{equation}%
the three complex structures are then can be given by 
\begin{align}
\mathbf{J}_{1}& =e^{0}\wedge e^{1}-e^{2}\wedge e^{3}=\frac{1}{4}d\left(
r^{2}\sigma _{1}\right) ,  \notag \\
\mathbf{J}_{2}& =e^{0}\wedge e^{2}+e^{1}\wedge e^{3}=\frac{1}{4}d\left(
r^{2}\sigma _{2}\right) ,  \notag \\
\mathbf{J}_{3}& =e^{0}\wedge e^{3}-e^{1}\wedge e^{2}=\frac{1}{4}d\left(
r^{2}\sigma _{3}\right) .
\end{align}%
For solutions with neutral flat base space, we can express the
four-dimensional base metric in terms of the forms 
\begin{equation}
\sigma _{1}^{\prime }=\sin \phi d\theta -\cos \phi \sinh \theta d\psi ,\text{
\ \ \ \ }\sigma _{2}^{\prime }=\cos \phi d\theta +\sinh \theta \sin \phi
d\psi ,\text{ \ \ \ \ }\sigma _{3}^{\prime }=d\phi +\cosh \theta d\psi 
\end{equation}%
or 
\begin{equation}
\chi _{1}^{\prime }=-\sin \psi d\theta +\cos \psi \sinh \theta d\phi ,\text{
\ \ \ \ }\chi _{2}^{\prime }=\cos \psi d\theta +\sinh \theta \sin \psi d\phi
,\text{ \ \ \ \ }\chi _{3}^{\prime }=d\psi +\cosh \theta d\phi 
\end{equation}%
satisfying 
\begin{equation}
d\sigma _{i}^{\prime }=-\frac{1}{2}f_{ijk}\sigma _{j}^{\prime }\wedge \sigma
_{k}^{\prime },\text{ \ \ \ \ }d\chi _{i}^{\prime }=\frac{1}{2}f_{ijk}\chi
_{j}^{\prime }\wedge \chi _{k}^{\prime }
\end{equation}%
where $f_{ijk}$ are the structure constants of $SO(2,1).$ The metric takes
the form 
\begin{eqnarray}
ds_{4}^{2} &=&dr^{2}+\frac{r^{2}}{4}\left( -\sigma _{1}^{\prime 2}-\sigma
_{2}^{\prime 2}+\sigma _{3}^{\prime 2}\right)   \notag \\
&=&dr^{2}+\frac{r^{2}}{4}\left( -\chi _{1}^{\prime 2}-\chi _{2}^{\prime
2}+\chi _{3}^{\prime 2}\right)   \notag \\
&=&dr^{2}+\frac{r^{2}}{4}\left( -d\theta ^{2}-\sinh ^{2}\theta d\phi
^{2}+\left( d\psi +\cosh \theta d\phi \right) ^{2}\right) .
\end{eqnarray}%
In this case the three two-forms satisfying the hypersymplectic algebra can
be given by 
\begin{equation}
\mathbf{J}_{i}=\frac{1}{4}d\left( r^{2}\sigma _{i}^{\prime }\right) ,\text{
\ \ \ \ }i=1,2,3.
\end{equation}%
As an example, we consider the STU\ model with space-time signatures ($2,3)$
and $(3,2)$ described by the prepotential $\mathcal{V=}X^{1}X^{2}X^{3}$. The
solutions with signatures ($1,4)$ were considered in \cite{sc, unique}. The
metric is given by the general form 
\begin{equation}
ds_{5}^{2}=\kappa ^{2}f^{4}\left( d\tau +w\right) ^{2}+f^{-2}\left[ dr^{2}+%
\frac{r^{2}}{4}\left( -d\theta ^{2}-\sinh ^{2}\theta d\phi ^{2}+\left( d\psi
+\cosh \theta d\phi \right) ^{2}\right) \right] 
\end{equation}%
where $\kappa ^{2}=1$ for solutions with ($2,3)$ signature and $\kappa
^{2}=-1$ for solutions with ($3,2)$ signature. We consider the simple case
with vanishing $\Psi ^{I}$ in (\ref{ge}) which implies that $f^{-2}X_{I}$
are given in terms of harmonic functions $H_{I}$ on the base space and we
obtain 
\begin{eqnarray}
f^{-6} &=&H_{1}H_{2}H_{3}  \notag \\
X^{1} &=&\left( \frac{H_{3}H_{2}}{H_{1}^{2}}\right) ^{1/3},\text{ \ \ \ }%
X^{2}=\left( \frac{H_{3}H_{1}}{H_{2}^{2}}\right) ^{1/3},\text{ \ \ }%
X^{3}=\left( \frac{H_{2}H_{1}}{H_{3}^{2}}\right) ^{1/3}.
\end{eqnarray}%
The gauge fields are given by 
\begin{equation}
F^{I}=d\left( X^{I}f^{2}\left( d\tau +w\right) \right) .
\end{equation}%
As $G_{+}=0,$ we obtain from (\ref{s}) that $dw$ is anti-self-dual and we
can set 
\begin{equation}
w=\frac{J}{r^{2}}\left( d\phi +\cosh \theta d\psi \right) 
\end{equation}%
with a constant $J$.

One can also consider solutions with neutral base given by an analytic
continuation of the Eguchi-Hanson metric given by \cite{bar} 
\begin{equation}
ds_{4}^{2}=W^{-1}dr^{2}+\frac{r^{2}}{4}\left( -\sigma _{1}^{\prime 2}-\sigma
_{2}^{\prime 2}+W\sigma _{3}^{\prime 2}\right) 
\end{equation}%
with 
\begin{equation*}
W=1-\frac{a^{4}}{r^{4}}.
\end{equation*}%
In this case, the hypersymplectic structure is defined by 
\begin{equation}
\mathbf{J}_{i}=d\left( \frac{1}{4}r^{2}W^{1/2}\sigma _{i}^{\prime }\right) .
\end{equation}%
One can also have analytic continuations of the general hyper-K\"{a}hler $N$%
-multi-centered Gibbons-Hawking metrics which admit tri-holomorphic Killing
vector field \cite{gh, th} and obtain hypersymplectic metrics. Recall that
these metrics are described by 
\begin{eqnarray}
ds^{2} &=&V^{-1}(dx^{4}+\theta )+V\left( \left( dx^{1}\right) ^{2}+\left(
dx^{2}\right) ^{2}+\left( dx^{3}\right) ^{2}\right)   \notag \\
\nabla \times \theta  &=&\nabla V,\text{ \ \ \ \ \ \ \ \ \ \ \ \ \ \ \ \ \ \
\ }V=\eta +\sum_{i=1}^{N}\frac{\varrho }{|x-x_{i}|}  \label{gm}
\end{eqnarray}%
where $\eta $ and $\varrho $ are constants. The tri-holomorphic Killing
vector is $\partial _{x^{4}}$ and $\theta $ $=\theta _{i}dx^{i}$. For $\eta
=0$ and $N=1,$ we  obtain flat space and for $\eta =0$ and $N=1,$ we obtain
Eguchi-Hanson metric. One can analytically continue the metrics (\ref{gm})
and obtain hypersymplectic metrics. For example we can consider the metrics 
\begin{equation}
ds^{2}=V^{-1}(dx^{4}+\theta )+V\left( -\left( dx^{1}\right) ^{2}-\left(
dx^{2}\right) ^{2}+\left( dx^{3}\right) ^{2}\right)   \notag
\end{equation}%
with the hypersymplectic structure given by 
\begin{align}
\mathbf{J}_{1}& =(dx^{4}+\theta )\wedge dx^{1}-Vdx^{2}\wedge dx^{3},  \notag
\\
\mathbf{J}_{2}& =(dx^{4}+\theta )\wedge dx^{2}-Vdx^{3}\wedge dx^{1},  \notag
\\
\mathbf{J}_{3}& =(dx^{4}+\theta )\wedge dx^{3}+Vdx^{1}\wedge dx^{2}.
\end{align}%
All the solutions considered in \cite{original} which included
generalizations of BMPV\ black hole solutions \cite{bmpv}, rotating
Eguchi-Hanson and Taub-NUT solutions and solutions with Gibbons-Hawking base
space can be analytically continued to obtain solutions with neutral bases.
The continued hypersymplectic manifold will inherit the Killing fields of
the Euclidean metric \cite{bar}. \ However it must be emphasized that
neutral manifolds are less rigid than Riemannian manifolds. For example,
Killing vectors with zero norms can not exist in the Riemannian case. Not
all neutral hypersymplectic metrics can be obtained by analytic
continuations.

In general, hypersymplectic metrics can be written in terms of one function
in the form 
\begin{equation}
ds^{2}=\frac{\partial ^{2}Y}{\partial x\partial w}dxdw+\frac{\partial ^{2}Y}{%
\partial y\partial z}dydz+\frac{\partial ^{2}Y}{\partial y\partial w}dydw+%
\frac{\partial ^{2}Y}{\partial x\partial z}dxdz  \label{pl1}
\end{equation}%
where the function $Y$ satisfies the so-called the first Heavenly equation 
\cite{si, ple}%
\begin{equation}
\frac{\partial ^{2}Y}{\partial x\partial w}\frac{\partial ^{2}Y}{\partial
y\partial z}-\frac{\partial ^{2}Y}{\partial y\partial w}\frac{\partial ^{2}Y%
}{\partial x\partial z}=1.  \label{h1}
\end{equation}%
An alternative representation of hypersymplectic metrics is given by 
\begin{equation}
ds^{2}=dy\left( dw-\frac{\partial ^{2}S}{\partial x^{2}}dy-\frac{\partial
^{2}S}{\partial w\partial x}dz\right) -dz\left( dx+\frac{\partial ^{2}S}{%
\partial w^{2}}dz+\frac{\partial ^{2}S}{\partial w\partial x}dy\right) 
\label{pl2}
\end{equation}%
with $S$ satisfying the so-called second Heavenly equation \cite{ple} 
\begin{equation}
\frac{\partial ^{2}S}{\partial w\partial y}-\frac{\partial ^{2}S}{\partial
z\partial x}+\frac{\partial ^{2}S}{\partial w^{2}}\frac{\partial ^{2}S}{%
\partial x^{2}}-\left( \frac{\partial ^{2}S}{\partial x\partial w}\right)
^{2}=0.  \label{h2}
\end{equation}%
Many interesting four-dimensional hypersymplectic metrics with various types
of Killing vectors such as null Killing vectors and conformal Killing
vectors have been constructed (see for example \cite{ms, mw}). Using the
Heavenly equation formalism, a notable example of a class of non-compact
metrics on the cotangent bundles of Riemann surfaces with genus $\geq $ 1
was constructed in \cite{ov}.

In what follows we shall consider the $(2,2)$ analogs of pp-waves \cite{ple}
which in the notation of \cite{ms} take the form 
\begin{equation}
ds^{2}=dy\left( dw-Q(x,y)dy\right) -dzdx  \label{wav}
\end{equation}%
where $Q$ is an arbitrary function. These metrics have a null Killing vector 
$\partial _{w}$ which can be thought of as a neutral signature version of a
tri-holomorphic Killing vector \cite{mw}. The metrics (\ref{wav}) were also
considered in the context of twistors \cite{ward} and have also appeared in
the analysis of \cite{bar} and in the classification of neutral solutions
admitting Killing spinors \cite{sg}. Using our formalism we rewrite (\ref%
{wav}) in the form%
\begin{equation}
ds^{2}=2\left( \mathbf{E}^{1}\mathbf{E}^{\bar{1}}-\mathbf{E}^{2}\mathbf{E}^{%
\bar{2}}\right) 
\end{equation}%
with 
\begin{equation}
\mathbf{E}^{1}=\frac{1}{2\sqrt{2}}\left[ dw+\left( 1-Q\right) dy+i\left(
dz-dx\right) \right] ,\text{ \ \ \ \ \ \ }\mathbf{E}^{2}=\frac{1}{2\sqrt{2}}%
\left[ dw-\left( 1+Q\right) dy+i\left( dz+dx\right) \right] .
\end{equation}%
Then the hypersymplectic structures is expressed in terms of 
\begin{align}
\mathbf{J}_{1}& =\frac{1}{2}\left( dy\wedge dw-dz\wedge dx\right) ,  \notag
\\
\mathbf{J}_{2}& =\frac{1}{2}\left( dy\wedge dz+\left( dw-Qdy\right) \wedge
dx\right) ,  \notag \\
\mathbf{J}_{3}& =\frac{1}{2}\left( dy\wedge dz-\left( dw-Qdy\right) \wedge
dx\right) .
\end{align}%
Again as an example we again consider solutions of the STU\ model with $%
G_{+}=0.$ In this case we obtain 
\begin{eqnarray}
f^{-6} &=&\mathcal{H}_{1}\mathcal{H}_{2}\mathcal{H}_{3}  \notag \\
X^{1} &=&\left( \frac{\mathcal{H}_{2}\mathcal{H}_{3}}{\mathcal{H}_{1}^{2}}%
\right) ^{1/3},\text{ \ \ \ }X^{2}=\left( \frac{\mathcal{H}_{3}\mathcal{H}%
_{1}}{\mathcal{H}_{2}^{2}}\right) ^{1/3},\text{ \ \ }X^{3}=\left( \frac{%
\mathcal{H}_{2}\mathcal{H}_{1}}{\mathcal{H}_{3}^{2}}\right) ^{1/3}
\end{eqnarray}%
where $\mathcal{H}_{i}$ are harmonic functions on the base space described
by (\ref{wav}) which can be arbitrary functions of the coordinates $x$ and $%
y.$ As $dw$ is anti-self-dual we can for example set 
\begin{equation}
dw=\left( dy\wedge dw-dz\wedge dx\right) .
\end{equation}

In this paper we have considered a class of solutions admitting Killing
spinors of five dimensional ungauged supergravity with Abelian vector
multiplets. The base space of solutions with space-time signatures $(1,4)$, $%
(4,1)$, $(5,0)$ and $(0,5)$ are given in terms of hyper-K\"{a}hler
manifolds. The solutions of the five dimensional theories with space-time
signatures ($2,3)$ and $(3,2),$ the base manifold admits a hypersymplectic
structure \cite{hitch}.

Hypersymplectic geometry has a very rich structure and not all
hypersymplectic manifolds can be obtained from hyper-K\"{a}hler manifolds
via analytic continuation. All the examples considered in \cite{original,
unique} can be analytically continued to obtain solutions with
hypersymplectic base manifold. It would be of interest to construct many
explicit solutions and generalise our results to gauged five-dimensional
supergravity theories. We hope to report on this in a future publication

\bigskip

\textbf{Acknowledgements}: The work is supported in part by the National
Science Foundation under grant number PHY-1620505. The author would like to
thank M. Dunajski and J. Gutowski for useful discussions.

\end{document}